\documentstyle[a4wide,12pt]{article}
\begin{document}

\author{B. Linet \thanks{E-mail: linet@celfi.phys.univ-tours.fr}\\
\small Laboratoire de Math\'ematiques et Physique Th\'eorique \\
\small CNRS/UPRES-A 6083, Universit\'e Fran\c{c}ois Rabelais \\
\small Parc de Grandmont 37200 TOURS, France}
\title{\bf Entropy bound of a charged object and \\
electrostatic self-energy in black holes}
\date{}
\maketitle
\thispagestyle{empty}

\begin{abstract}

Without pretending to any rigour, we find a general expression of the 
electrostatic self-energy in static black holes with spherical symmetry. 
We determine the entropy bound of a charged object by assuming the existence 
of thermodynamics for these black holes. By combining these two 
results, we show that the entropy bound does not depend on the considered 
black hole.

\end{abstract}

\section{Introduction}  

By arguing from the generalized second law of thermodynamics for the Schwarzschild
black hole, Bekenstein \cite{bek1} conjectured in 1981 the existence of an upper bound on
the entropy of any neutral object. This derivation was immediately criticized
by Unruh and Wald \cite{unr} which point out that quantum effects produce a buoyany force
on the box containing the matter since this one is accelerated and, as a 
consequence, the generalized entropy always increases even if the entropy bound
is not verified. Recently, Bekenstein and Mayo \cite{bek2}, 
Hod \cite{hod} and then Linet \cite{lin} derived an upper
bound on the entropy of any charged object, initially found by Zaslavskii 
\cite{zas} in another context, by requiring the validity of thermodynamics of 
Reissner-Nordstr\"{o}m or Kerr-Newman black holes linearized with respect to 
the electric charge. In this proof, it is essential to take into account the electrostatic 
self-energy of the charged object in the Schwarzschild or Kerr black holes. 
As in the neutral case, Shimomura and Mukohyama \cite{shi} criticized this derivation
for the same reasons.

In this paper, we are not going to discuss these criticisms but we give understanding
of the relation between the entropy bound of a charged object, as obtained in the
original method of Bekenstein, and the electrostatic self-energy in 
static black holes with spherical symmetry. We suppose that the electromagnetic 
field is a test field minimally coupled to the metric. Without pretending to any rigour, 
we find a general expression of the electrostatic self-energy in these black holes.
Then, we assume the existence of thermodynamics for such black holes. In the neutral case, 
we obtain immediately the upper bound on the entropy of the object. In the charged case, 
we obtain an entropy bound which is independent on the choice of the black hole 
thanks to the found expression of the electrostatic self-energy.

The plan of the work is as follows. In section 2, we recall basic definitions about
the static black holes with spherical symmetry. We study the electrostatics in these
black holes in section 3. The purpose of this section is to conjecture the general
expression of the electrostatic self-energy. The derivation of the entropy bound
is fulfilled in section 4. We add in section 5 some concluding remarks. 

\section{Static black holes with spherical symmetry}

We consider a static black hole with spherical symmetry which contains for
example a $U(1)$-field  $A^{(Q)}_{\mu}$ with charge $Q$ and eventually other 
fields having no charge. 
The spacetime is asymptotically flat. The black hole is characterized by its mass 
$m$ and for instance the parameter $Q$. In the notations of Visser \cite{vis1}, 
there exists a coordinate system $(t,r,\theta ,\varphi )$ in which the metric can be 
written as
\begin{equation}\label{1}
ds^2=-{\rm e}^{-2\phi (r)} \left( 1-\frac{b(r)}{r}\right) dt^2+
\left( 1-\frac{b(r)}{r}\right)^{-1}dr^2+r^2d\theta^2+r^2\sin^2\theta d
\varphi^2
\end{equation}
where $b$ and $\phi$ are two functions of the radial coordinate $r$ with the 
condition $\phi (\infty )=0$. The horizon $r=r_H$ is defined by the equation 
$b(r_H)=r_H$ with the sssumptions that $\phi$ and its derivative are finite 
at $r=r_H$. So, metric (\ref{1}) is valid for $r>r_H$. The
quantity $b(\infty )/2$ is the mass $m$ of the black hole.

Since there exists a Killing horizon ${\cal H}$ at $r=r_H$, we have 
the generalization of the Smarr formula \cite{bar}
\begin{equation}\label{3}
m=\frac{\kappa {\cal A}}{4\pi}-\frac{1}{4\pi}\int_{\Sigma}dS_{\mu}
R^{\mu}_{\nu}\xi^{\nu}
\end{equation}
where $\xi^{\nu}$ is the timelike Killing vector and $\Sigma$ a spacelike hypersurface, 
which starts from ${\cal H}$ to reach the spatial infinity, with the surface element 
$dS_{\mu}$. ${\cal A}$ is the area $4\pi r_{H}^{2}$ of the horizon
${\cal H}$. The quantity $\kappa$ is the surface gravity of the horizon defined by
$\xi^{\alpha}\nabla_{\alpha}\xi^{\beta}=\kappa \xi^{\beta}$ at $r=r_H$.
For metric (\ref{1}), this latter definition  gives the expression
\begin{equation}\label{5}
\kappa =
\frac{{\rm e}^{-\phi (r_H)}}{2r_H}\left( 1-\frac{db}{dr}(r_H)\right)
\end{equation}
which would be a function of $m$ and $Q$. It will hereafter be assumed that 
$b'(r_H)<1$ to get $\kappa >0$.

For a black hole in any gravitational field theory, it is possible in general 
to define a function $S_{bh}$ of $m$ and $Q$ which satisfies the first law of 
the black hole mechanics \cite{wal,vis2,jac1,iye,jac2}. 
Discarding a variation of $Q$, we limit ourselves to consider the first law
\begin{equation}\label{10}
dm=\frac{\kappa}{2\pi}dS_{bh} \quad {\rm when}\quad dQ=0 \, .
\end{equation}
Within the Euclidean approach in quantum field theory at finite temperature
\cite{gib}, it is clear that $\kappa /2\pi$ is the Hawking temperature. 
So, $S_{bh}$ is well interpreted as the entropy of the black hole. In general, 
$S_{bh}\not= {\cal A}/4$.
In the system containing a black hole and matter with entropy $S$, we assume
the generalized second law of thermodynamics 
\begin{equation}\label{12}
dS_{bh}+dS\geq 0 \, .
\end{equation}

Metric (\ref{1}) can support an electrostatic test field with spherical
symmetry corresponding to an electric charge $q$ inside the horizon such that
$\mid q\mid \ll m$ and $\mid q\mid \ll \mid Q\mid$. The electromagnetic 
potential $A_{\mu}$, distinct from $A_{\mu}^{(Q)}$, has only a non-vanishing
component $A_{0}$ which satisfies the electrostatic equation 
$\partial_r(r^2{\rm e}^{\phi (r)}\partial_rA_{0})=0$ for $r>r_H$.
From this and the Gauss theorem, we obtain the solution
\begin{equation}\label{2a}
A_{0}(r)=qa(r) \quad {\rm with}\quad 
a(r)=\int_{r}^{\infty}\frac{{\rm e}^{-\phi (r)}dr}{r^2}
\end{equation}
whose electric field has a regular behaviour at the horizon. We denote 
$a_H=a(r_H)$.

We admit that a charged black hole with parameters $m$, $Q$ and electric
charge $q$ exists since the electrostatic test field $qa$ is regular at the 
horizon of metric (\ref{1}).
The variation of the mass $dm$ of the solution to another solution
resulting from a small amount $dq$ is given by
\begin{equation}\label{10e}
dm=\frac{\kappa}{2\pi}dS_{bh}+qa_Hdq\quad {\rm when}\quad dQ=0
\end{equation}
where $qa_Hdq$ representes the electrostatic energy at the horizon 
of the charge $dq$ in the exterior field $qa$.
We can rewrite first law (\ref{10e}) under the useful form
\begin{equation}\label{11}
dS_{bh}=\frac{2\pi}{\kappa}dm -\frac{2\pi}{\kappa}qa_Hdq \quad {\rm when}
\quad dQ=0 \, .
\end{equation}

We now give the example of the Reissner-Nordstr\"{o}m black hole for a
$U(1)$-field $A_{\mu}^{(Q)}$ minimally coupled. It is described by metric 
(\ref{1}) with
\begin{equation}\label{1b}
b^{(RN)}(r)=2M-\frac{Q^2}{r}\quad {\rm and}\quad 
\phi^{(RN)}(r)=0 \, .
\end{equation}
Its mass $m$ is $M$.
The horizon is located at $r_H=r_+=M+\sqrt{M^2-Q^2}$. The entropy $S_{bh}$
is $4\pi r_{H}^{2}$ and the surface gravity is
\begin{equation}\label{5b}
\kappa^{(RN)}=
\frac{\sqrt{M^2-Q^2}}{(M+\sqrt{M^2-Q^2})^2} \, .
\end{equation}
The electrostatic potential $a$ is $1/r$ and consequently $a_H=1/r_H$.

\section{Electrostatics in black holes}

We consider an electric test charge $e$ held fixed in metric (\ref{1}), 
satisfying $\mid e\mid \ll m$ and $\mid e\mid \ll \mid Q\mid$. Since $e$ and $Q$
are two different types of local charge, we can linearly determine
the electrostatic field generated by the charge $e$
without taking into account the backreaction. For a point charge $e$
located at $r=r_0$, $\theta =\theta_0$ and $\varphi =\varphi_0$, the 
electrostatic potential $V$ obeys the electrostatic equation
\begin{eqnarray}\label{13} 
\nonumber & & \frac{\partial}{\partial r}\left( r^2{\rm e}^{\phi (r)}
\frac{\partial}{\partial r} V
\right)+{\rm e}^{\phi (r)}\left( 1-\frac{b(r)}{r}\right)^{-1}
\left[ \frac{1}{\sin \theta}
\frac{\partial}{\partial \theta}\left( \sin \theta 
\frac{\partial}{\partial \theta}V\right) +\frac{1}{\sin^2\theta}\frac{\partial^2}{\partial \varphi^2} V\right] \\
& & =-4\pi e \delta (r-r_0)\delta (\cos \theta -\cos \theta_0)
\delta (\varphi -\varphi ) \, .
\end{eqnarray}
The physical solution to equation (\ref{13}) must have a regular behaviour at the
horizon and no electric flux through the sphere of radius $r_H$. 
The electric flux ${\cal F}_R[V]$ for $R>r_H$ has to have the value 
\begin{equation}\label{14}
{\cal F}_R[V]=-\int_{r=R}r^2{\rm e}^{\phi (r)}\partial_rV
\sin \theta d\theta d\varphi = 
\left\{ \begin{array}{ll}
4\pi e & R>r_0 \\
0 & R<r_0 \, . \end{array}\right. 
\end{equation}
Moreover, we must add the exterior potential $qa$ to the potential $V$.

The solution to equation (\ref{13}) is expressed as the sum of the elementary
solution in the Hadamard sense, denoted $V_C$, and a homogeneous solution. 
By virtue of the spherical symmetry of metric (\ref{1}), we induce that 
the homogeneous term is proportional to $a$. Moreover, the operator in 
equation (\ref{13}) is self-adjoint and consequently the expression of $V$ is 
symmetric in $r$ and $r_0$. Thus, $V$ has necessarily the form
\begin{equation}\label{15}
V=V_C(r,r_0,\theta ,\theta_0,\varphi ,\varphi_0)+sea(r_0)a(r)  \, .
\end{equation}
The potentiel $V_C$ has the Coulombian form at the neigbourhood of the point 
charge $e$. The parameter $s$, having the dimension of length, is to be determined in 
function of $m$ and $Q$ by taking into account the global condition (\ref{14}).

We now turn to the example of the Reissner-Nordstr\"{o}m
metric characterized by $M$ and $Q$. The electrostatic potential 
$V$ generated by an electric charge $e$ at $r=r_0$ and $\theta =0$ has already 
determined in closed form \cite{lea}
\begin{equation}\label{15b}
V^{(RN)}(r,\theta ,\varphi )=V_{C}^{(RN)}(r,\theta ,\varphi )+\frac{eM}{rr_0}
\end{equation}
where $V_{C}^{(RN)}$ has the expression
\begin{equation}\label{15a}
V_{C}^{(RN)}=\frac{e}{rr_0}\frac{(r-M)(r_0-M)-(M^2-Q^2)\cos \theta}
{[(r-M)^2+(r_0-M)^2-2(r-M)(r_0-M)\cos \theta -(M^2-Q^2)\sin^2\theta ]^{1/2}}
\end{equation}
having a singularity at $r=r_0$ and $\theta =0$. It should be a multiple of the
elementary solution derived in isotropic coordinates by Copson \cite{cop}.
The electric flux ${\cal F}_{r_H}[V_{C}^{(RN)}]$ through the sphere of radius $r_H$ is  
\begin{equation}\label{14b}
{\cal F}_{r_H}[V_{C}^{(RN)}] = -4\pi e\frac{M}{r_0}
\end{equation}
in accordance with the introduction of the homogeneous term into solution (\ref{15b}).
In the Reissner-Nordstr\"{o}m black hole, we see that $s=M$.

In general, it is not possible to find an exact solution to equation (\ref{13}).
However, we are going to consider two cases in which we can determine the parameter
$s$ appearing in form (\ref{15}).

\subsection{Analysis near of the horizon in the case $\phi =0$}

In the case of metric (\ref{1}) with $\phi =0$, the electrostatic equation 
(\ref{13}) can be written near the horizon as
\begin{eqnarray}\label{13a}
\nonumber & &\frac{\partial}{\partial r}\left( r^2
\frac{\partial}{\partial r}V\right) +\frac{r_H}{(1-b'(r_H))(r-r_H)}\left[
\frac{1}{\sin \theta}\frac{\partial}{\partial \theta}\left( \sin \theta 
\frac{\partial}{\partial \theta}V\right)
\frac{1}{\sin^{2}\theta}\frac{\partial^2}{\partial \varphi^2}V \right] \\
& & \approx -4\pi e \delta (r-r_0)\delta (\cos \theta -\cos \theta_0)
\delta (\varphi -\varphi_0)
\end{eqnarray}
for $r$ and $r_0$ near $r_H$. In the other hand, the electrostatic equation 
in a Reissner-Nordstr\"{o}m black hole is
\begin{eqnarray}\label{30}
\nonumber & & \frac{\partial}{\partial r}\left( r^2\frac{\partial}{\partial r} V^{(RN)}\right)
+\frac{r^2}{(r-r_-)(r-r_+)}\left[ \frac{1}{\sin \theta}\frac{\partial}{\partial \theta}
\left( \sin \theta \frac{\partial}{\partial \theta}V^{(RN)}\right) 
+\frac{1}{\sin^2\theta}\frac{\partial^2}{\partial \varphi^2}V^{(RN)}\right] \\
& & = -4\pi e \delta (r-r_0)\delta (\cos \theta -\cos \theta_0)
\delta (\varphi -\varphi_0) \, .
\end{eqnarray}
Near $r=r_+$, equation (\ref{30}) coincides with equation (\ref{13a})
if the parameters $M$ and $Q$ of the Reissner-Nordstr\"{o}m metric
are such that
\begin{equation}\label{9}
r_H=r_+(M,Q) \quad {\rm and}\quad 1-\frac{db}{dr}(r_H)=
\frac{2\sqrt{M^2-Q^2}}{r_+(M,Q)} \, .
\end{equation}
The elementary solution in the Hadamard sense is uniquely determined
in this neighbourhood and therefore we allow $V_C \approx V_{C}^{(RN)}$ 
for $r$ and $r_0$ near $r_H$ to be true. Hence from result (\ref{14b}), we deduce
\begin{equation}\label{14c}
{\cal F}_{r_H}[V_C]\approx -4\pi e\frac{M}{r_0} 
\end{equation}
where $M$ will be determined from (\ref{9}).
In the limit where $r_0\rightarrow r_H$, we get 
\begin{equation}\label{14d}
{\cal F}_{r_H}[V_C]=-4\pi e \left[ 1-\frac{1}{2}\left( 1-
\frac{db}{dr}(r_H)\right) \right] \, .
\end{equation}

We are now in a position to determine the parameter $s$ in form (\ref{15})
for a black hole with $\phi =0$ since $V$ must further satisfy the global 
condition (\ref{14}). Taking into account the electric flux (\ref{14d}),  
we obtain $s$ under the form
\begin{equation}\label{7}
s=\frac{1}{a_H}\left( 1-\kappa r_H \right) 
\end{equation}
where $\kappa$ is given by (\ref{5}) with $\phi =0$. We notice that $a_H=1/r_H$.

\subsection{A particular black hole with $\phi \not= 0$}

It is desirable to adopt a metric which allows us to find an exact solution 
to the electrostatic equation (\ref{13}) in the case $\phi \not= 0$, 
without referring to any gravitational theory of gravity. A  trivial example 
is a metric which is conformal to a Reissner-Nordstrom metric. Since
equation (\ref{13}), $\kappa$ and $a_H$ are conformally invariant 
then we find again formula (\ref{7}). 

In order to study a non-trivial case, we put the following metric 
\begin{eqnarray}\label{1a}
\nonumber & & ds^2=-\left( 1-\frac{\alpha}{R}\right)^{-2}
\left( 1-\frac{2m}{R}\right)dt^2+\left( 1-\frac{2m}{R}\right)^{-1}dR^2 \\
& & +R^2\left( 1-\frac{\alpha}{R}\right) \left(d\theta^2+
\sin^2\theta d\varphi^2\right)
\end{eqnarray}
in the coordinate system $(t,R,\theta ,\varphi )$. We take $0<\alpha <2m$.
Metric (\ref{1a}) is defined for $R>2m$ and $R=2m$ is a horizon.
So, metric (\ref{1a}) describes a black hole. 
We can easily go to form (\ref{1}) of the metric by performing the
change of radial coordinate $r=\sqrt{R(R-\alpha)}$ and we verify that 
$\phi \not= 0$.

The electrostatic potential $V^{(P)}$ due to an electric charge $e$ located at 
$(R_0,\theta_0,\varphi_0)$ is governed by the equation
\begin{eqnarray}\label{13b}
\nonumber & & \frac{\partial}{\partial R}\left[ (R-\alpha )^2
\frac{\partial}{\partial R} V^{(P)}\right] +
\frac{R-\alpha}{R-2m}\left[ \frac{1}{\sin \theta}
\frac{\partial}{\partial \theta}\left( \sin \theta 
\frac{\partial}{\partial \theta}V^{(P)}\right) +\frac{1}{\sin^2\theta}
\frac{\partial^2}{\partial \varphi^2} V^{(P)}\right] \\
& & = -4\pi e\delta (r-r_0)\delta (\cos \theta -\cos \theta_0)
\delta (\varphi -\varphi_0) \, .
\end{eqnarray}
We define a function X by setting $V^{(P)}=XR/(R-\alpha)$. From (\ref{13b}), 
we obtain 
\begin{eqnarray}\label{13c}
\nonumber & & \frac{\partial^2}{\partial R^2}X+\frac{2}{R}
\frac{\partial}{\partial R} X+\frac{1}
{(R-\alpha )(R-2m)} \left[ \frac{1}{\sin \theta}
\frac{\partial}{\partial \theta}\left( \sin \theta 
\frac{\partial}{\partial \theta}X\right) +\frac{1}{\sin^2\theta}
\frac{\partial^2}{\partial \varphi^2}X\right] \\
& & =-\frac{4\pi e}{R_0(R_0-\alpha)}
\delta (r-r_0)\delta (\cos \theta -\cos \theta_0)\delta (\varphi -\varphi_0)
\end{eqnarray}
which coincides with equation (\ref{30}).
So, the elementary solution in the Hadamard sense to equation (\ref{13b}) is
\begin{equation}\label{15c}
V_{C}^{(P)}=\frac{R}{R-\alpha}\frac{R_0}{R_0-\alpha}V_{C}^{(RN)}(R,\theta ,\varphi )
\end{equation}\label{9b}
for the parameters $M$ and $Q$ of the Reissner-Nordstr\"{o}m background such that
\begin{equation}\label{31}
2m+\alpha =2M\quad {\rm and}\quad 2m\alpha =Q^2 \, .
\end{equation}
The electric flux at the infinity of $V_{C}^{(P)}$ can be calculated from 
expression (\ref{15a}) of $V_{C}^{(RN)}$
\begin{equation}\label{8}
{\cal F}_{\infty}[V_{C}^{(P)}]=4\pi e \frac{R_0-M}{R_0-\alpha} \, .
\end{equation}
In the case of metric (\ref{1a}),
the electrostatic potential $a$ given by (\ref{2a}) has the expression
$1/(R-\alpha )$. By substracting $4\pi e$ from (\ref{8}), we find thereby 
$s=M-\alpha$. With the value of $M$ deduced from (\ref{31}), we therefore
have the expression of the parameter $s$ 
\begin{equation}\label{7b}
s=m-\frac{\alpha}{2} \, .
\end{equation}
The surface gravity $\kappa$ can be calculated by using the radial coordinate 
$R$ and we find
\begin{equation}\label{5c}
\kappa^{(P)} =\frac{1}{2(2m-\alpha)} \, .
\end{equation}
We have $a_H=1/(2m-\alpha )$ and so $\kappa^{(P)}/a_H=1/2$. 

From general result (\ref{7}) for black holes in the case 
$\phi =0$ and this particular case (\ref{7b}) with $\phi \not= 0$, 
we conjecture that the parameter $s$ can be expressed in terms of 
$a_H$ and $\kappa$ by the formula
\begin{equation}\label{6} 
s=\frac{1}{a_H}\left( 1-\frac{\kappa}{a_H}\right) \, .
\end{equation}

\section{Entropy bound of a charged object}

We now consider a charged object with a mass $\epsilon$, an
electric charge $e$ and a radius $\ell$ located at the position 
$(r_0,\theta_0,\varphi_0)$ in metric (\ref{1}) characterized by the parameters
$m$ and $Q$. We suppose that this object has its own gravitational
field negligeable, i.e. $\epsilon \ll m$ and $\epsilon \ll \mid Q\mid$ 
and that its electric field satisfies the conditions of validity for 
a test field mentioned in the previous section.
The energy ${\cal E}$ of the charged object is the sum of the energy
obtained by integrating on the energy-momentum of matter and the
energy of the electromagnetic field.

The energy of the electrostatic field $V+qa$ in the background metric 
is given by
\begin{eqnarray}\label{16}
\nonumber & & W_{em}=-\frac{q}{4\pi}
\int\sqrt{-g}g^{00}g^{ij}\partial_iV\partial_jadrd\theta d\varphi \\
& &-\frac{1}{8\pi}\int\sqrt{-g}g^{00}g^{ij}\partial_iV\partial_jV
drd\theta d\varphi 
\end{eqnarray}
which $V$ is governed by (\ref{13}) with the global condition (\ref{14}).
The first term in expression (\ref{16}) is the electrostatic energy $W_{elect}$
of the charge $e$ in the exterior field $qa$. By performing an integration by part,
we obtain
\begin{equation}\label{17}
W_{elect}=qea(r_0)
\end{equation}
since the integral on the sphere of radius $r_H$ vanishes according to 
(\ref{14}). The second term in expression (\ref{16}) is infinite for a point charge. The divergence has
a Coulombian type resulting from the part $V_C$ of the potential $V$  
expressed in form (\ref{15}). It should be incorporated at the mass of the charge. Nevertheless, it remains a 
finite part from the homogeneous term which leads to the 
electrostatic self-energy $W_{self}$
\begin{equation}\label{18}
W_{self}=\frac{1}{2}e^2s[a(r_0)]^2 \, .
\end{equation}

With formulas (\ref{17}) and (\ref{18}), we obtain
\begin{equation}\label{19}
{\cal E}=\epsilon \sqrt{-g_{00}(r_0)}+qea(r_0)+\frac{1}{2}e^2s[a(r_0)]^2 \, .
\end{equation}
The last state in which the charged object is just outside the horizon 
is defined by the position $r_0$ which is related to $\ell$ by the formula
\begin{equation}\label{20}
\ell \sim 2(r_0-r_H)^{1/2}r_{H}^{1/2}\frac{1}{[1-b'(r_H)]^{1/2}}
\end{equation}
by assuming that the proper radial distance $\ell$ is very small. Its energy 
${\cal E}_{last}$ has expression (\ref{19}) calculated for $r_0$ determined by 
(\ref{20}). We find thereby
\begin{equation}\label{21}
{\cal E}_{last}\sim \kappa \epsilon \ell +qea_H+\frac{1}{2}se^2a_{H}^{2} \, .
\end{equation}

We are now in a position to use the original method of Bekenstein for finding
the entropy bound for the entropy $S$ of this object. 
We consider in fact this charged object in the black holes defined by
metric (\ref{1}), characterized by $m$ and $Q$, plus the electrostatic
test field $qa$, characterized by the electric charge $q$. To obtain the expression 
of $S_{bh}(m,Q,q)$ of the charged black hole 
linearized with respect to $q^2$, we perform 
the integration of the first law (\ref{11}) under the form
\begin{equation}\label{22a}
S_{bh}(m,Q,q)\approx \overline{S}_{bh}(m,Q)-\frac{\pi}{\kappa}a_Hq^2 \, .
\end{equation}
The generalized entropy  of the state just ouside the horizon is 
$S_{bh}(m,Q,q)+S$. When the charged object falls in the horizon, the final 
state is a black hole with the new parameters
\begin{equation}\label{21a}
m_f=m+{\cal E}_{last} \quad q_f=q+e \quad {\rm and}\quad Q_f=Q \, .
\end{equation}
However, the entropy reduces to $S_{bh}(m_f,Q_f,q_f)$ in this final state. 
Now, we write down the generalized second law (\ref{12}) of thermodynamics 
\begin{equation}\label{22}
S_{bh}(m_f,Q_f,q_f)\geq S_{bh}(m,Q,q)+S \, .
\end{equation}
The increase of entropy linear in ${\cal E}_{last}$ can be calculated 
from the first law (\ref{11}). However, we want to keep the terms in $e^2$
and this is why we use the linearized expression (\ref{22a}). We obtain 
\begin{equation}\label{23}
dS_{bh}=\frac{\pi}{\kappa}\left( 2{\cal E}_{last}-e^2a_H-2eqa_H \right) \, .
\end{equation}
By inserting (\ref{21}) into (\ref{23}), we thus have
\begin{equation}\label{24}
dS_{bh}=2\pi \epsilon \ell +\pi e^2(sa_H-1)a_H\frac{1}{\kappa} \, .
\end{equation}
With the help of (\ref{24}), the generalized second law (\ref{22}) gives
the desired entropy bound of the charged object
\begin{equation}\label{25}
S\leq 2\pi \epsilon \ell +\pi e^2(sa_H-1)a_H\frac{1}{\kappa}
\end{equation}
which seems to be dependent on the parameters of the considered black hole.

With our conjecture (\ref{6}) on the value of $s$, we obtain the 
entropy bound
\begin{equation}\label{26}
S\leq 2\pi \left( \epsilon \ell -\frac{1}{2}e^2\right)
\end{equation} 
already derived from thermodynamics of the Schwarzschild black hole.

\section{Conclusion}

We have assumed the existence of thermodynamics of static black holes 
with spherical symmetry but fortunately without having has to know the 
expression of the entropy $S_{bh}$ in terms of the parameters of
the black holes. Then, we have found the upper bound (\ref{25}) on the 
entropy of a charged object which is dependent on the expression of the 
electrostatic self-energy.  In the neutral case, it reduces immediately 
to the usual entropy bound \cite{bek1}.

The crucial point is the determination of the electrostatic self-energy
(\ref{18}), i.e. the value of the parameter $s$ appearing in form
(\ref{15}). We have obtained very strong indications that $s$ can
be expressed in terms of the surface gravity $\kappa$ and
the value $a_H$ of the electrostatic potential at the horizon. We have found
formula (\ref{6}):
\[
s=\frac{1}{a_H}\left( 1-\frac{\kappa}{a_H}\right) \, .
\]
A rigorous proof of this result is beyond our scope. 
By admitting this result, we then obtain an entropy bound for a charged object
where the parameters of the black hole disappear
as in the Schwarzschild case \cite{bek2,hod}.


\newpage

\begin{thebibliography}{99}
\bibitem{bek1} Bekenstein, J.D. (1981). {\em Phys. Rev. D} {\bf 23}, 287.
\bibitem{unr} Unruh, W.G. and Wald, R.M. (1982). {\em Phys. Rev. D}
{\bf 25}, 942.
\bibitem{bek2} Bekenstein, J.D. and Mayo, A.E. (1999). gr-qc/9903002.
\bibitem{hod} Hod, S. (1999). gr-qc/9903011.
\bibitem{lin} Linet, B. (1999). {\em Gen. Rel. Grav.} {\bf 31}, 1609.
\bibitem{zas} Zaslavskii, O. (1992). {\em Gen. Rel. Grav.} {\bf 24}, 973.
\bibitem{shi} Shimomura, T. and Mukohyama, S. (1999). gr-qc/9906047.
\bibitem{vis1} Visser, M. (1992). {\em Phys. Rev. D} {\bf 46}, 2445.
\bibitem{bar} Bardeen, J.M., Carter, B. and Hawking, S.W. (1973).
Commun. Math. Phys. {\bf 31}, 161.
\bibitem{wal} Wald, R.M. (1993). {\em Phys. Rev. D} {\bf 48}, 3427.
\bibitem{vis2} Visser, M. (1993). {\em Phys. Rev. D} {\bf 48}, 5697.
\bibitem{jac1} Jacobson, T., Kang, G. and Myers, R.C. (1994).
{\em Phys. Rev. D} {\bf 49}, 6587.
\bibitem{iye} Iyer, V. and Wald, R.M. (1994). {\em Phys. Rev. D} {\bf 50}, 846.
\bibitem{jac2} Jacobson, T., Kang, G. and Myers, R.C. (1995).
{\em Phys. Rev. D} {\bf 52}, 3518.
\bibitem{gib} Gibbons, G.W. and Hawking, S.W. (1977). 
{\em Phys. Rev. D} {\bf 15}, 2752.
\bibitem{lea} L\'eaut\'e, B. and Linet, B. (1976). {\em Phys. Lett. A} 
{\bf 58}, 5.
\bibitem{cop} Copson, E.T. (1978). {\em Proc. Roy. Soc. Edinburgh A}
{\bf 80}, 201.

\end{thebibliography}
\end{document}